\documentclass{cimento}
\usepackage{graphicx}
\title{Comparison of cuprates and nuclear matter pairing properties. Quartet ($\alpha$-particle) condensation in nuclear systems.}

\author{P. Schuck\from{ins:x}}
\instlist{\inst{ins:x}Institut de Physique Nucl\'eaire, Orsay, France {\it and} Laboratoire de Physique et de Mod\'elisation des Milieux Condens\'es, Grenoble, France}

\begin{document}
\maketitle
\begin{abstract}
A comparison of pairing properties in cuprates and nuclear matter is briefly discussed. Quartet ($\alpha$-particle) condensation is a very important aspect of nuclear physics. The physics of the Hoyle state in $^{12}$C will be outlined and its crucial role for the existence of life on earth explained.
\end{abstract}

\section{Introduction}

In this contribution, I will treat two subjects: i) nuclear pairing in comparison with pairing in cuprates and ii) quartet ($\alpha$-particle) condensation in nuclear systems. Both condensation phenomena are extremely important in nuclear physics. Nuclear pairing shows multiple manifestations in nuclei, for instance a dramatic reduction of the moment of inertia for deformed nuclei with respect to its rigid body value. In neutron stars, one believes that the neutrons form a superfluid with a lattice of vortices due to the star's rotation. Neutron superfluidity will be contrasted with superconductivity in cuprates.\\
Alpha-clustering is, for instance, of outmost importance for element production in the universe, above all the $^{12}$C production. This will be discussed in the context of the Hoyle state in $^{12}$C.

\section{Cuprates vs nuclear matter}

\begin{figure}
  \includegraphics[width=6cm]{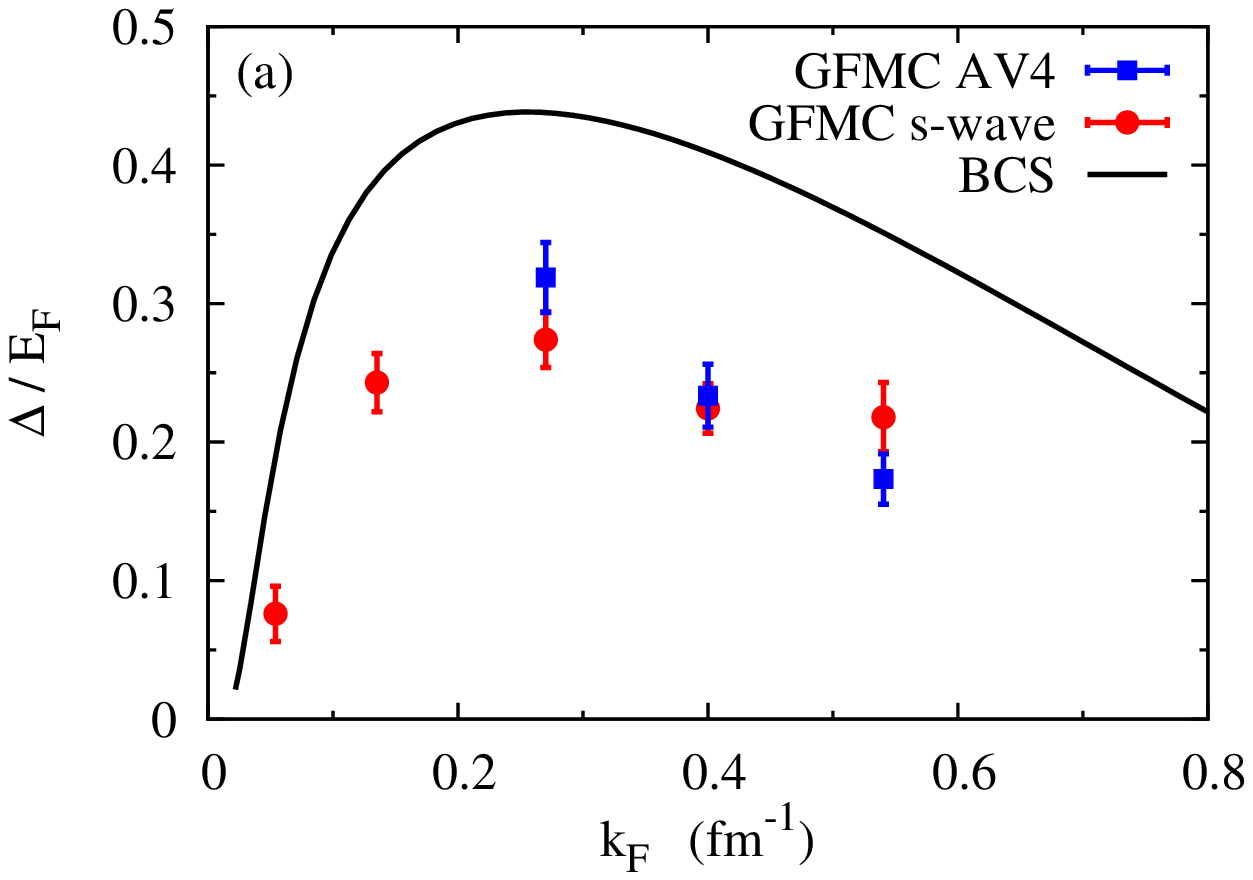}
  \includegraphics[width=6cm]{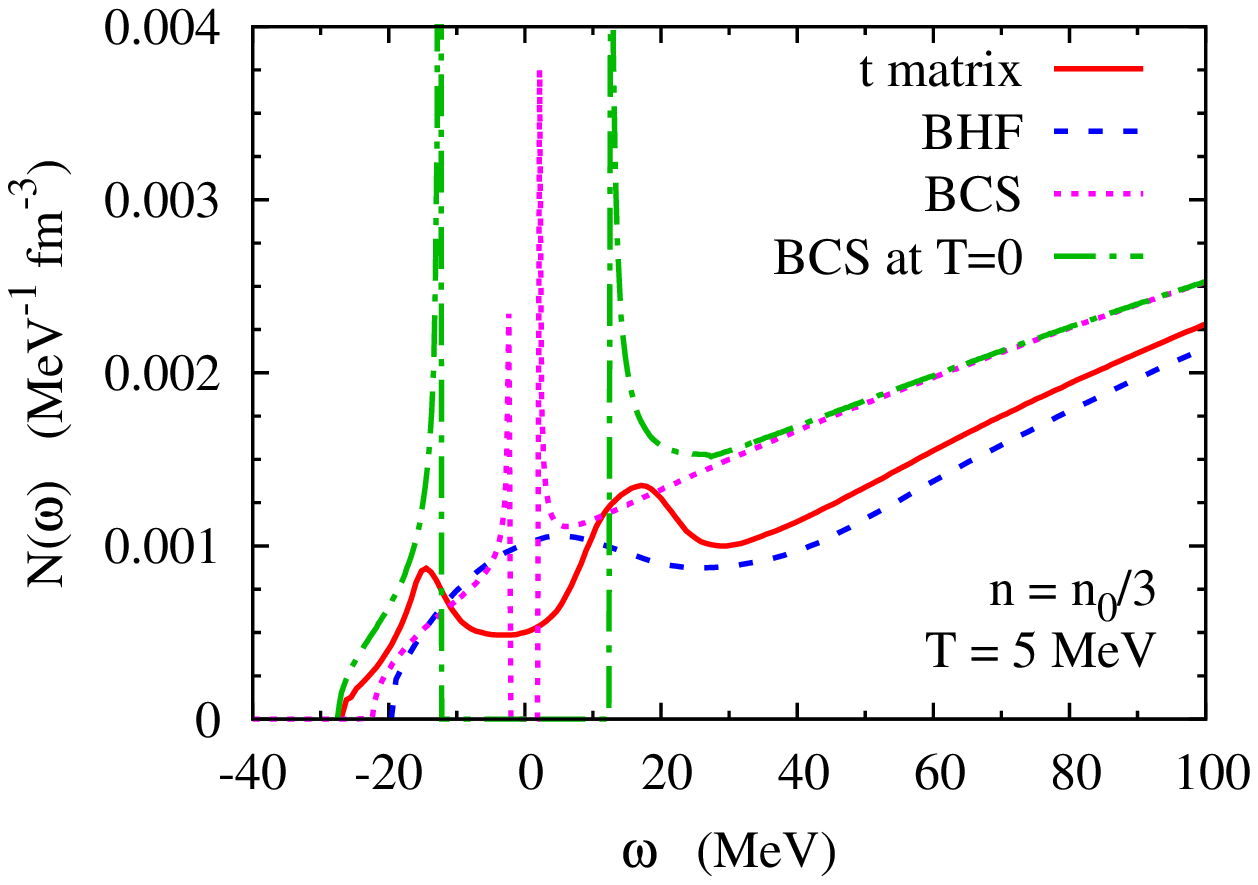}
  \caption{\label{pseudo} Left: the ratio gap ($\Delta $) over Fermi energy ($E_{\rm F}$) for neutron-neutron pairing is displayed as a function of Fermi momentum ($k_{\rm F}$) in neutron matter. Squares (blue) (full circles (red)) are from Green's function Monte Carlo (GFMC) calculations with slightly different ingredients. Figure from \cite{ref:PR}. Right:  Pseudo-gap in nuclear matter, full (red) line in the deuteron channel in nuclear matter. Disregard other lines. Figure from \cite{ref:PR}.}
\end{figure}

\begin{figure}
  \includegraphics[width=6cm]{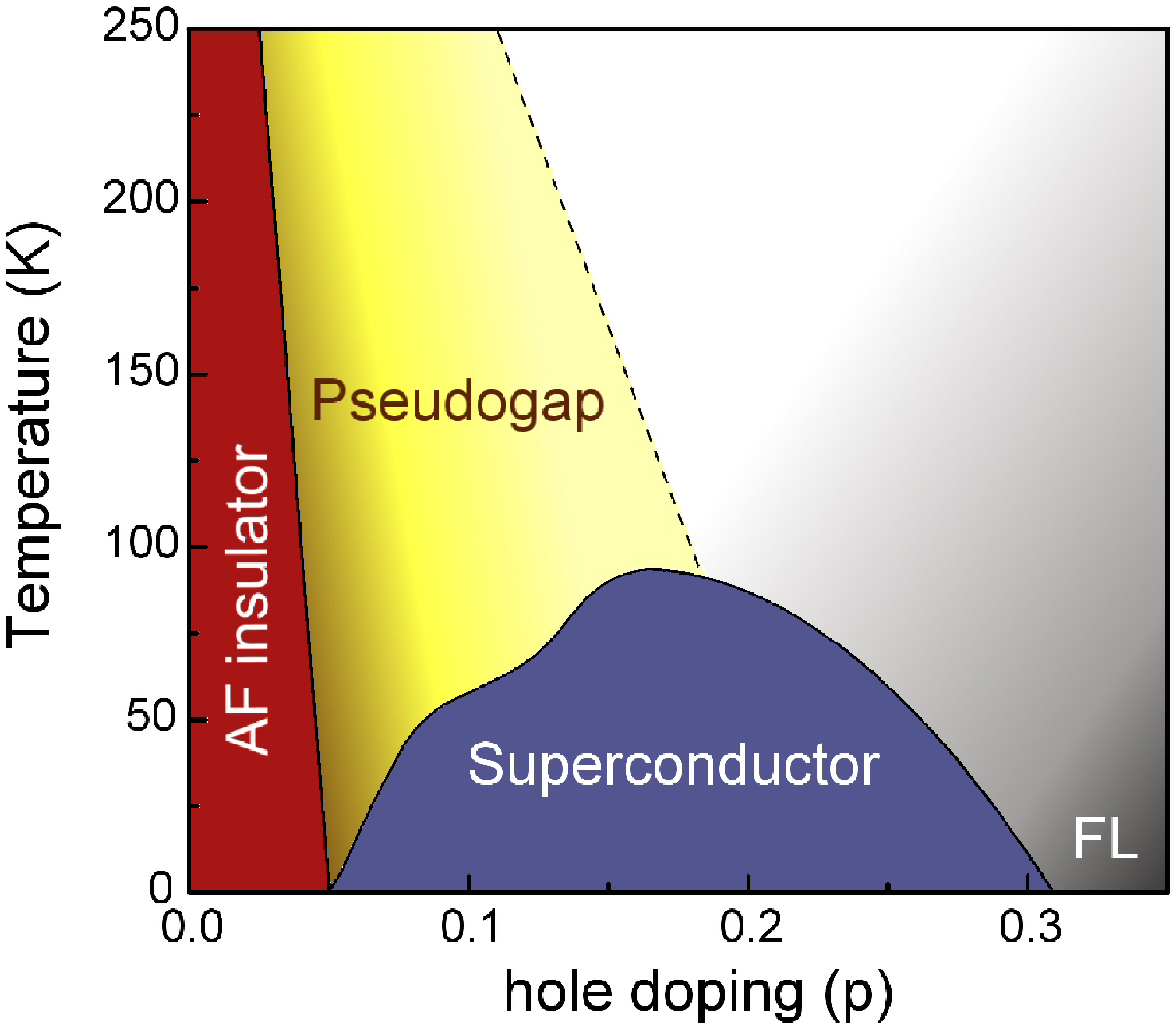}
  \includegraphics[width=6cm]{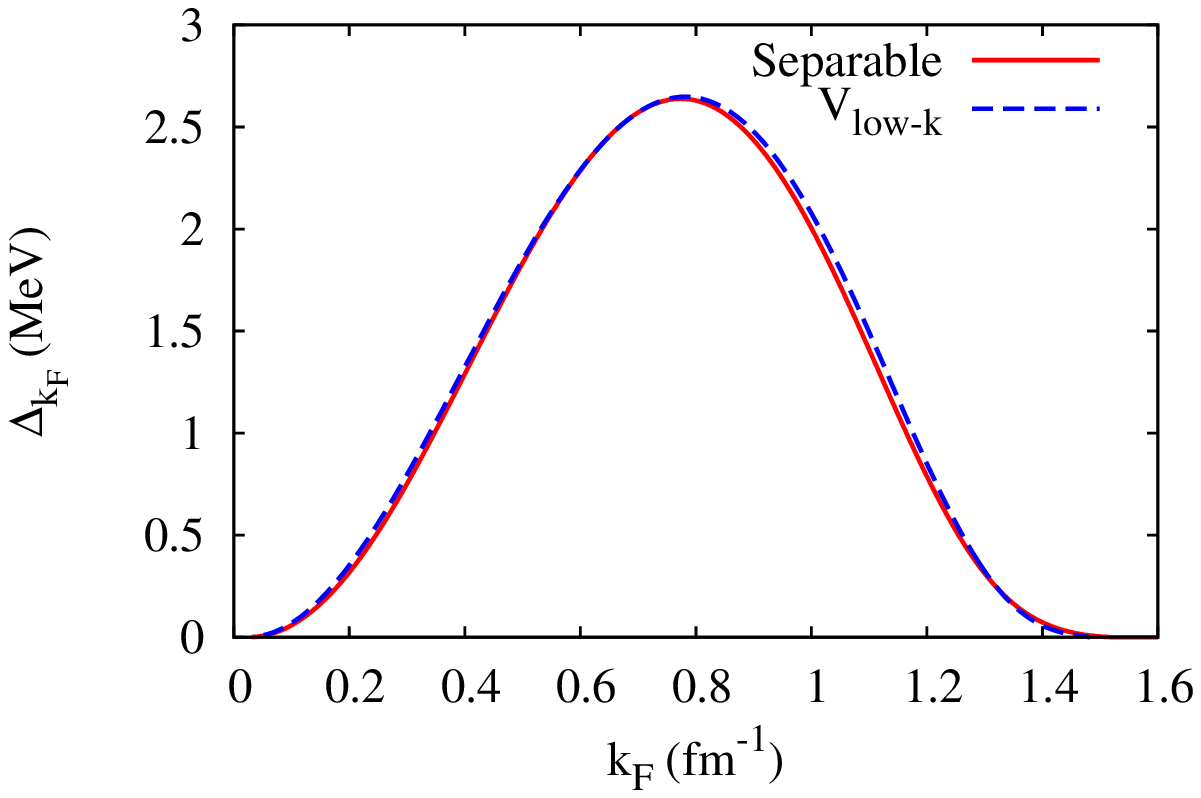}
  \caption{\label{highTc-phase}Left: schematic phase diagram of a high Tc superconductor with the superconducting dome limited by the critical temperature Tc as a function of the electron-hole density. Figure from \cite{ref:Cyril}, provided by C. Proust. Right: Superfluid ``dome'' of the nuclear gap obtained with with two effective pairing forces as a function of Fermi momentum \cite{ref:Urban}. Figure provided by M. Urban.}
\end{figure}

Pairing in nuclei and nuclear matter is much stronger than pairing of electrons in ordinary metals. For neutron-neutron pairing, this stems from the fact that two neutrons almost form a bound state in the spin singulet state while a proton and neutron bind to the deuteron in the spin triplet state. The ratio gap to Fermi-energy, $\Delta / E_{\rm F}$, can be as large as 1/5, see Fig.\ref{pseudo}, whereas in ordinary metals this ration is by orders of magnitude smaller.\\
Nuclear pairing, therefore, resembles more high Tc superconductivity than standard metallic one. There are, indeed, some similarities
between high Tc pairing and nuclear pairing other than the high ratio $\Delta / E_{\rm F}$. For example looking at a  phase diagram of a high Tc superconductor in Fig.2 where the critical temperature is shown as a function of doping, one sees that Tc follows the so-called dome shape, a much discussed subject in the condensed matter community. In nuclear matter there exists also some sort of dome for the gap as a fucntion of density or $k_{\rm F}$, see Fig.\ref{highTc-phase}. This dome-like shape in the nuclear case is simply a consequence of the fact that the nuclear (pairing) force is of finite range, that is, also in momentum, $k$-space, it is of finite range. As density ($k_{\rm F}$) tends to zero, so does the gap, since where is no matter, no gap exists. At high $k_{\rm F}$, the force tends to zero and also there the gap then goes to zero. Naturally a dome-like shape develops. The dome-shape of Tc in high Tc materials may eventually have the same origin. However, other explanations are certainly possible, see a recent publication \cite{ref:arXiv}. On the other hand in nuclear matter, as in high-Tc materials, also a pseudo-gap phase exists. In Fig.\ref{pseudo}, right panel, we show a typical pseudo-gap formation in the n-p spin triplet (deuteron) channel. In nuclear physics this pseudogap definitely is due to pair fluctuations. However, apparently in high Tc materials the pseudo gap phase is due to other processes \cite{ref:Antoine}.
In conclusion, the phase diagram of pairing in nuclear matter has, like for cuprates, a superfluid dome and a pseudo-gap phase. However, the underlying physics may be different. 

\section{Alpha-particle (quartet) condensation}

Alpha-particle condensation has become an important subject in nuclear cluster physics since the article by Tohsaki, Horiuchi, Schuck, R\"opke (THSR) appeared in 2001 \cite{ref:thsr} where the possibility of an $\alpha$-particle condensation in states close to the $\alpha$-particle disintegration threshold was considered for the first time for self-conjugate nuclei such as $^{12}$C and $^{16}$O. The considered $\alpha$ condensate wave function was given, e.g., for $^{12}$C, by

\begin{equation}
\Psi_{\mbox{THSR}} \propto {\mathcal A} \psi_1\psi_2\psi_3 \equiv {\mathcal A}|B\rangle 
\label{THSRwf}
\end{equation}
with
\[ \psi_i = e^{-(({\bf R}_i-{\bf X}_G)^2)/B^2}\phi_{\alpha_i}~;~~~~~~\phi_{\alpha_i} = e^{-\sum_{k<l}({\bf r}_{i,k}-{\bf r}_{i,l})^2/(8b^2)} \]



\begin{figure}
  \parbox{5.5cm}{\vspace{0cm}
    \includegraphics[height=4.5cm]{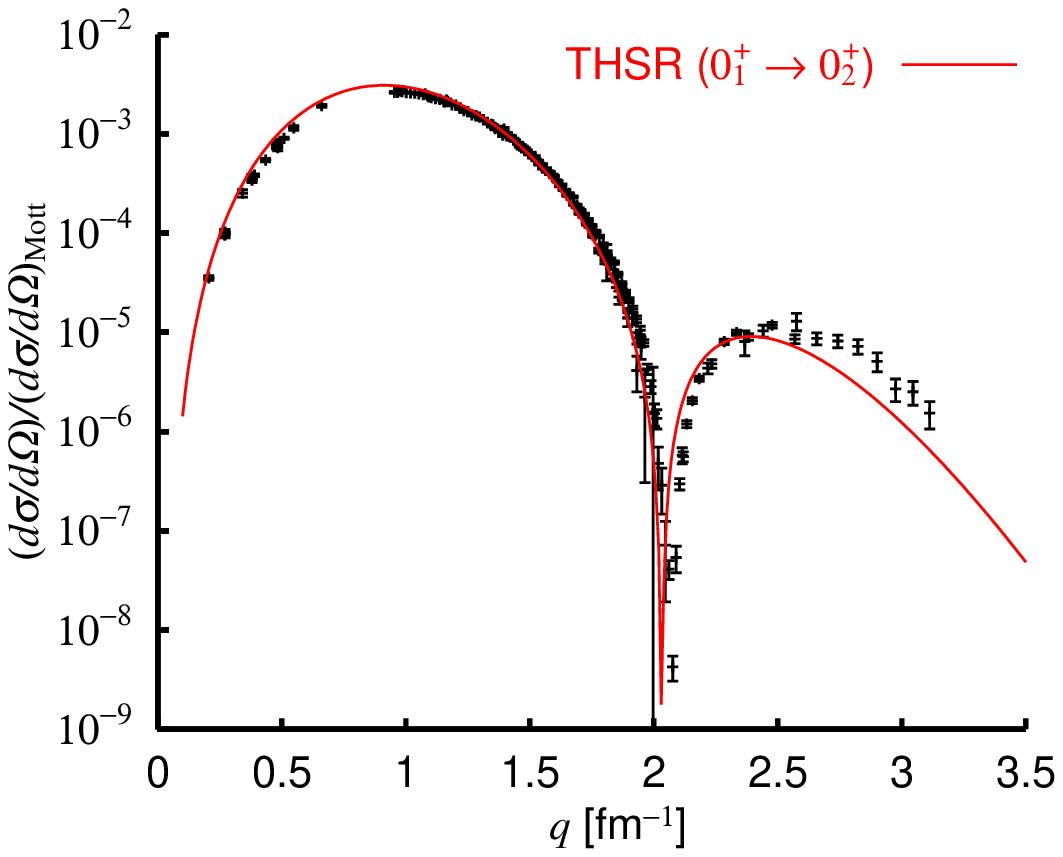}\hspace{0.5cm}}
    \parbox{5.5cm}{\vspace{0cm}
\includegraphics[height=4.5cm]{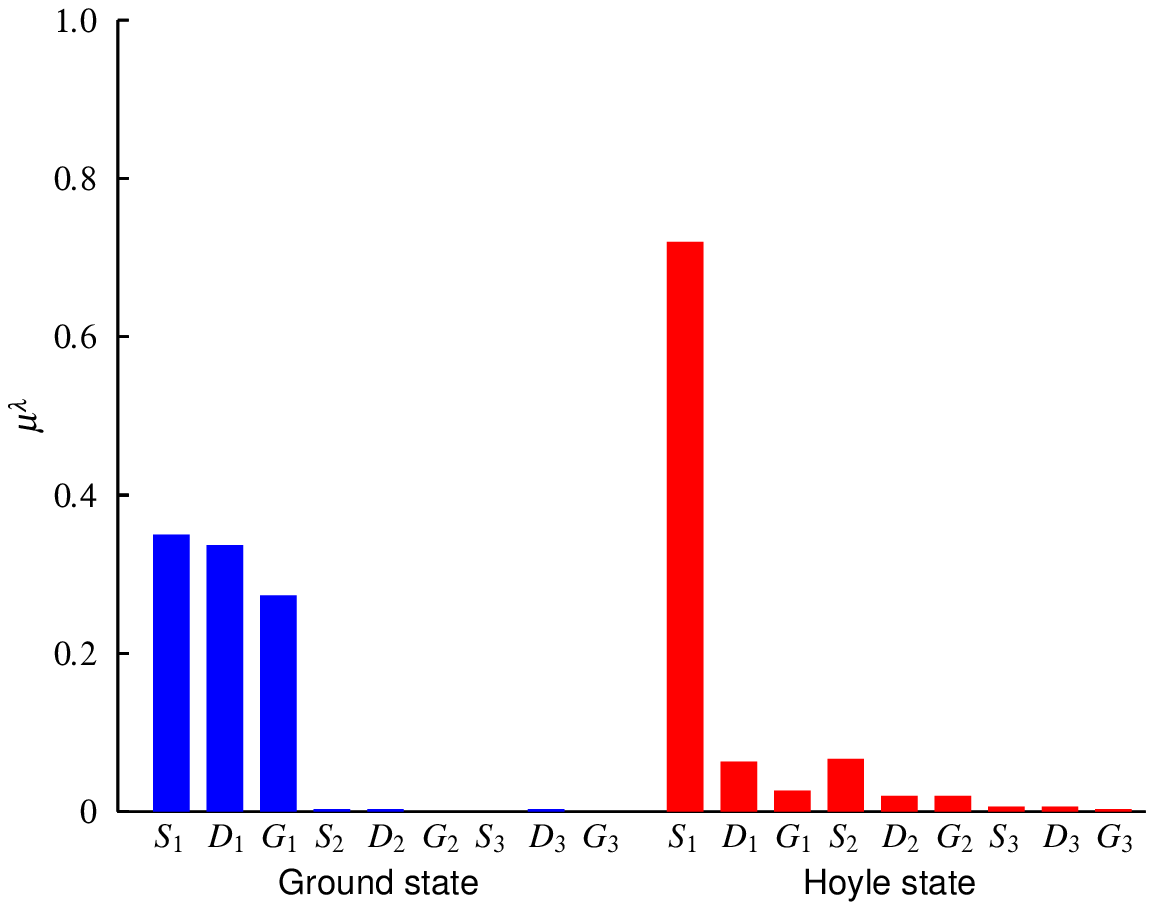}\hspace{0.5cm}}
\caption{\label{QMC}
Inelastic form factors, see, e.g., \cite{ref:RMP}, calculated from the THSR wave function, left panel. Right panel: The $\alpha$ occupation probabilities in the ground and Hoyle states, see \cite{ref:RMP}.}
\end{figure}

\noindent
In (\ref{THSRwf}) the ${\bf R}_i$ are the c.o.m. coordinates of $\alpha$ particle '$i$' and ${\bf X}_G$ is the total c.o.m. coordinate of $^{12}$C. ${\mathcal A}$ is the antisymmetrizer of the twelve nucleon wave function with $\phi_{\alpha_i}$ the intrinsic translational invariant wave function of the $\alpha$-particle '$i$'. The whole 12 nucleon wave function in (\ref{THSRwf}) is, therefore, translationally invariant. Please note that we suppressed the scalar spin-isospin part of the wave function. The special Gaussian form given in Eq.  (\ref{THSRwf}) was chosen in \cite{ref:thsr} to ease the variational calculation but it is known that for light nuclei a Gaussian ansatz for the wave function yields very good results. The condensate aspect lies in the fact that (\ref{THSRwf}) is a (antisymmetrized) product of three times the same $\alpha$-particle wave function and is, thus, analogous to a number projected BCS wave function in the case of pairing.  This twelve nucleon wave function has 
two variational parameters, $b$ and $B$. It possesses the remarkable property that for $B=b$ it is a pure harmonic oscillator Slater determinant whereas for $B \gg b$ the $\alpha$'s are at low density so far apart from one another that the antisymmetrizer can be dropped and, thus, (\ref{THSRwf}) becomes a simple product of three $\alpha$ particles, all in identical 0S states, that is, a pure condensate state, see discussion in \cite{ref:RMP}. The minimization of the energy with a Hamiltonian containing an effective nucleon-nucleon force determined 15 years earlier independently  allows to obtain a reasonable value for the ground state energy of $ ^{12}$C. Variation of energy under the condition that the state (\ref{THSRwf}) is orthogonal to the previously determined ground state allows to calculate the first excited $0^+$ state, i.e., the Hoyle state situated at 7.56 MeV. While the size of the individual $\alpha$ particles remains very close to their free space value ($b \simeq$ 1.37 fm), the variationally determined $B$ parameter takes on about three times this value.  The THSR approach reproduces very well all known experimental data of the Hoyle state. This concerns for instance the inelastic form factor, electromagnetic transition probability, and position of energy, see for more details \cite{ref:RMP}. The inelastic form factor is shown in  Fig.\ref{QMC}. We see indeed very good agreement with the experimental data. It should be mentioned that this agreement comes with no adjustable parameter and that the magnitude of the inelastic form factor is very sensitive to the radius of the Hoyle state. The radius of the Hoyle state is obtained at about 3.8 fm what is equivalent to a volme larger by a factor 3-4 in comparison with the one of the ground state (radius = 2.4 fm). This is indeed a very unusual nuclear state!\\
In the right panel of Fig.\ref{QMC} we show the single $\alpha$ occupancies of the $\alpha$'s in the ground state and in the Hoyle state. We see that as the occupancies are democratically distributed over several states in clear conformity with the SU3 limit of the shell model for the ground state, the Hoyle state shows an over 70\% occupancy of the $\alpha$'s in the 0S orbit, the other occupancies being down by over a factor of ten. This is a clear sign of condensation.\\
The Hoyle state is very important for the $^{12}$C production in the universe and, thus, for life on earth because it is the doorway of the so-called triple $\alpha$ reaction where in stars two $\alpha$'s first form the unstable $^8$Be nucleus (lifetime $\sim$ $10^{-17}$ s) and then during its lifetime a third $\alpha$ joins resonating with the 7.56 MeV state in $^{12}$C. Subsequently the Hoyle state decays by $\gamma$ emission to the ground state. The presence of the Hoyle state at the right energy accelerates the $^{12}$C production in the universe by several orders of magnitude. The Hoyle state is named after the astro-physicist Fred Hoyle who predicted this state at the right energy earlier to its discovery.\\
Besides the Hoyle state, one believes that Hoyle analog states are present in heavier self-conjugate nuclei like $^{16}$O, etc.

\acknowledgments
The author acknowledges close collaboration on the subject of $\alpha$-particle condensation with Y. Funaki, H. Horiuchi, G. R\"opke, A. Tohsaki, T. Yamada. With respect to nuclear pseudo-gap the collaboration with G. R\"opke is appreciated. Discussions with Antoine Georges concerning the physics of cuprates are greatfully acknowledged.

\end{document}